# Topological Corner Modes by Composite Wannier States in Glide-Symmetric Photonic Crystal

*Zhenzhen Liu[1,2,#], Xiaoxi Zho[3,4,#], Guochao Wei[5], Lei Gao[3,4], Bo Hou[6], and Jun-Jun Xiao[1,\*]*

[1]Shenzhen Engineering Laboratory of Aerospace Detection and Imaging, College of Electronic and Information Engineering, Harbin Institute of Technology (Shenzhen), Shenzhen 518055, China.

[2]College of Science, Shantou University, Shantou 515063, China.

[3]School of Optical and Electronic Information, Suzhou City University, Suzhou 215104, China

[4]School of Physical Science and Technology, Soochow University, Suzhou 215006, China

[5]School of Mathematical and Physical Sciences, Wuhan Textile University, Wuhan 430200, China

[6]Wave Functional Metamaterial Research Facility, The Hong Kong University of Science and Technology (Guangzhou), Guangzhou 511400, China

\* E-mail: eiexiao@hit.edu.cn

#These two authors contributed equally.

**Abstract:**

Second-order topological insulators can be characterized by their bulk polarization, which is believed to be intrinsically connected to the center of the Wannier function. In this study, we demonstrate the existence of second-order topological insulators that feature a pair of partially degenerate photonic bands. These arise from the nonsymmorphic glide symmetry in an all-dielectric photonic crystal. The center of the maximally localized Wannier function (MLWF) is consistently located at the origin but is not equivalent with respect to the sum of constituent polarizations. As a result, topological corner modes can be identified by the distinctly hybridized MLWFs that truncate at the sample boundary. Through full-wave numerical simulations paired with microwave experiments, the second-order topology is clearly confirmed and characterized. These topological corner states exhibit notably unique modal symmetries, which are made possible by the inversion of the Wannier bands. Our results provide an alternative approach to explore higher-order topological physics with significant potential for applications in integrated and quantum photonics.

## 1. Introduction

Topological physics has attracted great attentions and gained noteworthy developments over the past decades[1–2]. Though originally proposed and succeeded in condensed-matter physics, distinct topological phases have been extensively studied in various classical wave systems, e.g., photonic, phononic and elastic systems which offers great flexibility for exploration of various topological phase transitions and direct experimental observation on topologically localized states[3–7]. Despite of the rapid developments in both topological physics theory and experiential measurement, there are still many issues and new proposals that are underestimated. Most noticeably, higher-order topological states represent a striking example. Beyond the traditional bulk-edge correspondence, many types of higher-order topological insulators (HOTIs) have been found which host lower-dimensional boundary states that are believed to be topologically protected[8-13]. For example, second-order topological insulator has gapped one-dimensional (1D) edge states and zero-dimensional (0D) corner states in the band gap[14]. In terms of the highly localized topological corner states, photonic nano-cavity with extremely high Q-factor[15] and topological laser with low-threshold[16] have been designed and experimentally demonstrated.

The underlying theory of higher-order topological concepts have been evolved spirally in recent years. Rooted in modern electronic band theory, the concept of polarization in periodic quantum crystal is essential in describing solid state materials and is at the core of the understanding the topological phases of electronic condensed matters[17,18]. The bulk band polarization **P** was shown essentially equivalent to the Berry phase, and the energy bands with discontinuous (non-smooth varying) Bloch wavefunction lead to the quantization of the bulk polarization[19]. Regardless of the distinct thermodynamic statistic characteristics between electron and photon/phonon, this modern polarization theory for periodic crystals has also been applied to characterize the Bloch bands of neutral bosonic systems such as photonic crystal or phononic crystals, showing the effectiveness to predict band-structure topological effects such as stable surface modes[20-24]. The use of non-Abelian Berry phase to examine the presence topological nontrivial topological phase in photonic crystal has also been successful[25].

As a matter of fact, the bulk polarization encodes the 'averaged' position of the spatially-resolved Wannier functions, namely the so-called "Wannier centers", which further facilitates establishing the bulk-boundary-corner correspondence for the polarization components in various crystals[26]. Note that the Wannier band concept has been widely adopted and played a crucial role in studying topological photonic and phononic crystals, taking advantage of space group and eigenmode symmetry indicators[27,28]. The connection between the momentum-space topological indexes and the position-space Wannier functions provides new perspective and tools to identify topological physics associated phenomena[26]. Yet the explicit maximally localized Wannier functions (MLWF) were barely explored and not explicitly presented beyond electronic systems. In this regard, Gupta and Bradlyn recently showed the successful application of Wannier-function method for studying the topological phase transition and the topological defect states in one-dimensional photonic crystal[29]. The Wannier-function approach enables one to determine not just the symmetry indicators and Wannier centers for photonic bands, but also the full band representation spanned by the set of Wannier functions in position space. The full band representations by Wannier functions have inspired further insights for higher-dimensional system, as well as for multiple band systems. Particularly, for the 2D Su-Schrieffer-Heeger (SSH) model constructed in square lattice, the bulk polarization is connected to the 2D Zak phase via $\theta_i = 2\pi P_i$ for $i = x, y$. In this case, a nontrivial system has quantized bulk polarizations $(P_x, P_y) = (1/2, 1/2)$, i.e., the Wannier center is located at the corner of the primitive unit cell, which indicates a nontrivial phase with fractional corner charge of 1/4[28]. Then, the corner states emerge at the corner boundaries, featuring the higher-order topology explained by filling anomaly[18,19,23,28]. However, recent explorations of topological insulators have made it apparent that a bulk polarization representation is not always possible[30] such as in the case of type-III corner states[31]. Thus, whether the concept of Wannier bands can be extended beyond the current Wannier center picture is highly relevance and remains open in developing higher-order topological band theory.

In this study, we show that Wannier bands (defined for energy bands labeled by $m$ and $n$) featured by the combined bulk polarization components $(P_x^m, P_y^n)$ in nonsymmorphic insulators are physically meaningful. This is because the physical manifestations of spatially-resolved Wannier functions can be well-defined by the hybridization of the constituent Wannier functions. We further show that this leads to the emergence of different types of second-order topological phases, which host new types of corner states localized around the corner truncated by perfect electric conductor (PEC) guaranteed by effective 'charge' filling anomaly yielded by the obstruction of composite Wannier states. Additionally, we show that under the glide symmetry, the bound charges localized at the edge are fractionally quantized whereas the charges localized at the corner cells are different with regard to the boundary, which can also give rise to the corner states in finite-sized samples. We present our results in full-wave numerical simulations and experimentally confirm the theoretical and the numerical results in 2D photonic crystal in the microwave band.

## 2. Theoretical analysis and numerical results

Let us start the theory form a 2D crystalline insulators, for which the bulk polarization $\mathbf{P} = P_x \mathbf{a_x} + P_y \mathbf{a_y}$ where $\mathbf{a}_i$ ($i = x, y$) are primitive lattice vectors. The energy band $E_m(\mathbf{k})$ has polarization components $P_i = \oint d^2\mathbf{k} \text{Tr}[\mathcal{A}_i(\mathbf{k})]$, where $\mathcal{A}_i$ is the Berry connection with elements $[\mathcal{A}_i(\mathbf{k})]_{m,n} = i\langle u_m(\mathbf{k})| \partial_{k_i} |u_n(\mathbf{k})\rangle$, and $|u_m(\mathbf{k})\rangle = e^{-i\mathbf{k}\cdot\mathbf{r}}\psi_{n,k}$ is the periodic Bloch eigenstate of the occupied band $m$ at crystal momentum $\mathbf{k} = (k_x, k_y)$ [18,19]. To this end, all observed topological insulators fit into this framework of quantized bulk polarization, or mathematical generalizations thereof. The components of $\mathbf{P}$ along different directions can be rewritten as $P_i = \oint dk_j p_i(k_j)$ [30] with the $k_j$-sector polarization reading as

$$p_i(k_j) = \frac{i}{2\pi} \log \det \left[ e^{i \int_k^{k+2\pi} \mathcal{A}_i(\mathbf{k}) dk_i} \right] = \sum_{m=1}^{N} v_i^m(k_j) \text{ for } i, j = x, y \text{ and } i \neq j, \quad (1)$$

which is a summation over all the occupied bands $m$. The sets of Wannier centers $v_i^m$ along $x(y)$ direction as a function of $k_y(k_x)$ is the so-called hybrid Wannier bands, e.g., $v_i^m(k_j)$, intimately related to the Berry phase for a given $k_j$. However, the correlation (or the associated relation) between the two components $v_x$ and $v_y$ for orthogonal Cartesian axis is not well defined when multiple bands interact (entangled) or are partially degenerated. In such scenarios it is possible to construct different Berry connections while maintaining geometry symmetries by band exchanging. We stress that this represents the major contribution of this work which shows the possibility to make distinguishable topological phases with identical **P** summed over all bands below a complete gap.

To explore the combined effects of polarization components, we examine scenarios featuring two bands (schematically illustrated in Figure 1a) that are partially degenerate due to nonsymmorphic symmetry. In this context, we define $P_{x(y)} = P_{x(y)}^1 + P_{x(y)}^2$. It becomes evident that two distinct configurations of Wannier centers can represent the polarization components, namely $(P_x^1, P_y^1)$, $(P_x^2, P_y^2)$ and $(P_x^1, P_y^2)$, $(P_x^2, P_y^1)$. For example, for non-vanishing bulk polarization **P** = (1/2,1/2), the system could probably host two Wannier bands with centers equivalent to polarization (1/2,1/2), (0,0) or (1/2,0), (0,1/2), which have been revealed in our recent work[31]. Figures 1b and 1c schematically show one of the possible cases that the two $k$-sector Wannier bands remains similar for band-1 and band-2. However, Figures 1e and 1f are for the case where the two Wannier bands interchange with respect to $k_y(k_x)$ sector. Namely, the red and blue curves around 0 and 0.5 are inverted in Figure 1f with respect to those in Figure 1c. Fundamentally, these two scenarios can be represented schematically by the phase accumulation, or geometric phase, in two correlated orthogonal directions. To elucidate the origin more distinctly, Figures 1d and 1g depict two possible evolution schemes of eigenmodes on the two energy bands. Specifically, the polarizations for individual bands along XM and YM are identical, as determined by the Bloch mode pattern's parity at X/Y and M; However, the polarization along XM is opposite to that along YM, a consequence of the swapping of the eigenstates at points X/Y [18]. The insets in Figures 1d and 1g show the polarization vectors from the two bands with identical sum. It is noteworthy to remark that the constituent Wannier centers with equal sum exert significant effect on the intrinsic topological features, such as the fractional charge and the resultant corner states, as we shall show in the following.

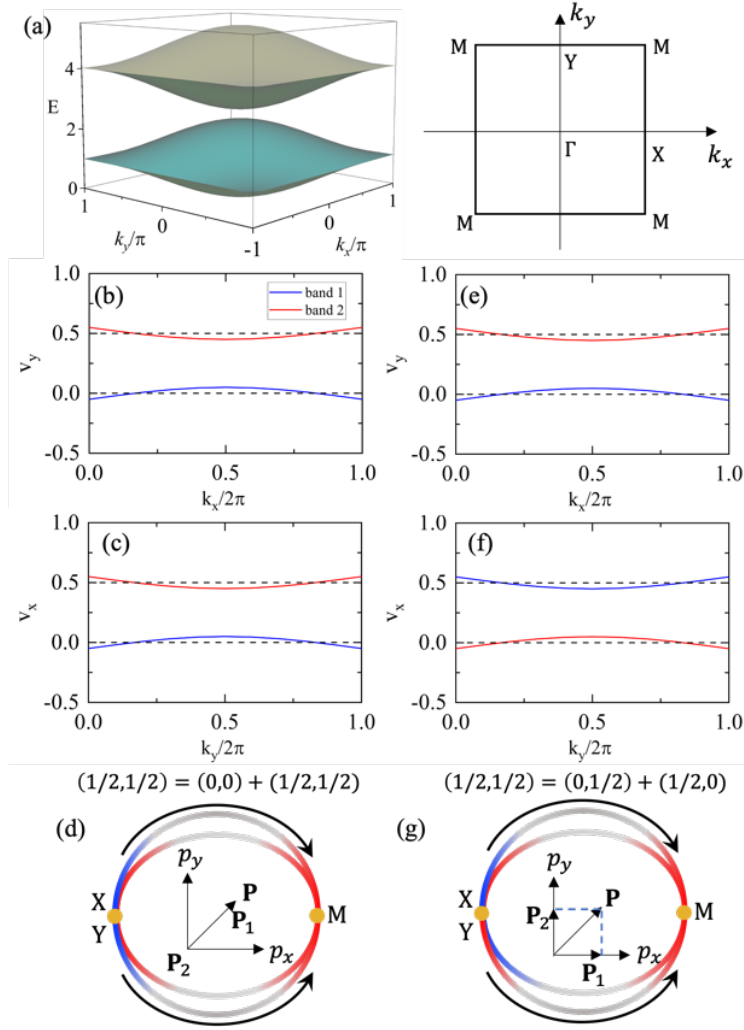

**Figure 1**. Schematic of the Wannier center position interchange with respect to two energy bands below a complete gap. (a) Two bands below a complete gap that are partially degenerated at the BZ boundary, guaranteed by crystal symmetry. (b)-(c) Wannier bands $v_{y,x}(k_{x,y})$ for the two bands numbered as 1 and 2. (d) The possible phase evolution of eigenmodes of the degenerate bands along XM and YM to mimic the Wannier bands combination $(1/2,1/2) = (0,0) + (1/2,1/2)$ in (b) and (c). (e)-(f) Another energy band-resolved Wannier bands $v_{y,x}(k_{x,y})$ with regard to the case in (b)-(c). (g) The possible phase evolution of eigenmodes of the degenerate bands to mimic the Wannier bands combination $(1/2,1/2) = (0,1/2) + (1/2,0)$ in (d) and (g). Here, the even and odd parity of the eigenstates at X and M are indicated by the red and blue colors, respectively.

As demonstrated in our previous study[31], the selection of a supercell plays a crucial role in determining the topological phases, especially in the case of total polarization (1/2, 1/2), which is ensured by both the glide symmetry and the mirror symmetry. In this context, Wannier bands assume a pivotal position in characterizing the topological phase transition. The corresponding band-resolved polarization components serve as a finer topological identifications and classifications that captures the observable and distinct corner states that arise from the domain interface. To explore the consequences of these two different combinations of Wannier bands in more concrete and realistic systems, we consider a finite, glide symmetric photonic crystal in square lattice, as shown in Figure 2a. The unit cell outlined by black dashed box consists of four dielectric rods placed in air. The rods are with radius $r$ and the lattice constant is $a$. The gray rods are marked as type-A with relative permittivity $\varepsilon_A$, while the orange ones are marked as type-B with permittivity $\varepsilon_B$. Note that setting $\varepsilon_{A,B} = 1$ indicates the absence of the cylinders in a sublattice (e.g., the gray or orange ones). At the first step, we suppose $\varepsilon_A = \varepsilon_B = 9.8$ and the system exhibits four-fold degeneracy at M point due to the band folding effect. Figure 2b shows the band structure for the four lowest energy bands (see more details on the symmetry analysis and the eigenmode patterns in Supporting Information S1).

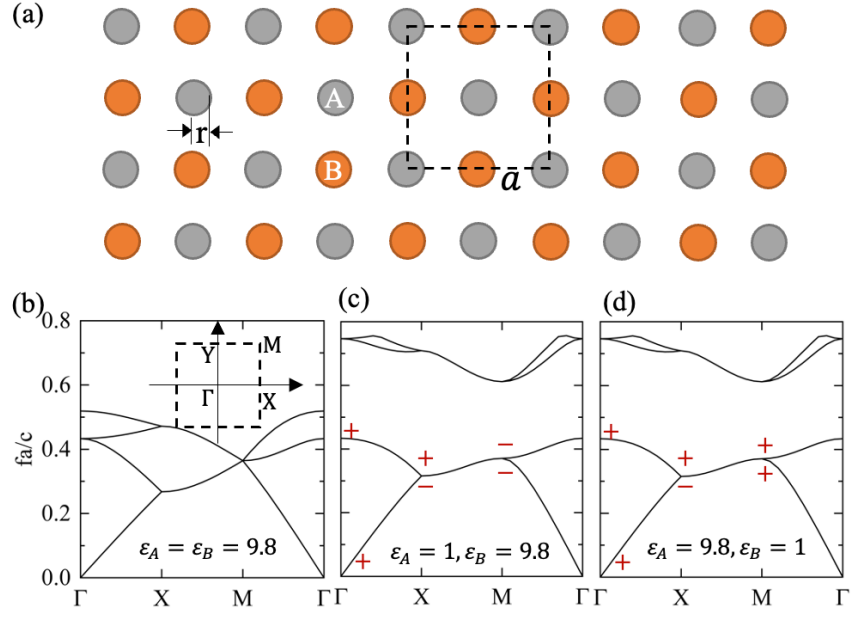

**Figure 2.** Schematic of the glide-symmetric photonic crystal and the TM band structure. (a) Schematic of the unit cells composed by subsites A, B. (b) The TM band structure for parameters with (b) $\varepsilon_A = \varepsilon_B = 9.8$, (c) $\varepsilon_A = 1$, $\varepsilon_B = 9.8$, and (d) $\varepsilon_A = 9.8$, $\varepsilon_B = 1$. The other parameters are set as $r = 0.143a$. The parities at high symmetric points are marked by "$\pm$". Inset in (b) shows the corresponding first Brillouin zone with high-symmetric points labeled by $\Gamma$, X (Y) and M.

By setting a discrepancy in $\varepsilon_A$ and $\varepsilon_B$, the system simultaneously obeys the glide symmetry $G_x: (x, y) \to (-x + a/2, y + a/2)$ and $G_y: (x, y) \to (x + a/2, -y + a/2)$, fourfold rotational symmetry $C_4: (x, y) \to (y, x)$, translation symmetry $\tau: (x, y) \to (x + a/2, y + a/2)$, inversion symmetry $I: (x, y) \to (-x, -y)$, mirror symmetry $M: (x, y) \to (-x, y)$, and time reversal symmetry $T$, respectively. The symmetries $G_x$ and $G_y$ are crucial to build the topological band gaps. With these symmetries, the fourfold degeneracy is split into a pair of double degenerated bands on the MX and MY lines because $\Theta_i^2 \psi_{n,k} = -\psi_{n,k}$, where $\Theta_i = G_i T$ ($i = x, y$). In particular, we have $\Theta_i I = -I\Theta_i$ at X(Y) point of the first Brillouin zone, whereas $\Theta_i I = I\Theta_i$ at M point. Therefore, the double degenerate eigenstates $\psi_{n,k}$ and $\Theta_i \psi_{n,k}$ have the opposite parities for the inversion $I$ at X and Y points, on the contrary, they have the identical parities for the inversion $I$ at M point (see Supporting Information S1). We can now define the symmetry-indicator invariant for a set of bands as $[\Pi_1^{(2)}] = \#\Pi_1^{(2)} - \#\Gamma_1^{(2)}$, where $\#\Pi_1^{(2)}$ is the number of states at momentum $\Pi$ with inversion eigenvalue $+1$. Therefore, the bulk polarization is $\left(\left[M_1^{(2)}\right] + \left[Y_1^{(2)}\right], \left[M_1^{(2)}\right] + \left[X_1^{(2)}\right]\right)/2 = (1/2, 1/2)$, a hallmark of higher-order topology[19].

Figures 2c and 2d show that for two specific cases of $\varepsilon_A \neq \varepsilon_B$, a gap opens around $f = 0.5a/c$ (here $c$ the velocity of light in vacuum) between the second and the third bands. We note that for a system with inversion symmetry, the calculation of **P** can be substantially simplified by checking the parities of eigenstates at the high-symmetry points in the BZ as follows[15,31]

$$P_i = \frac{1}{2}(\Sigma_m \ q_i^m \bmod 2), (-1)^{q_i^m} = \frac{\eta(X_m)}{\eta(\Gamma)}, \qquad (2)$$

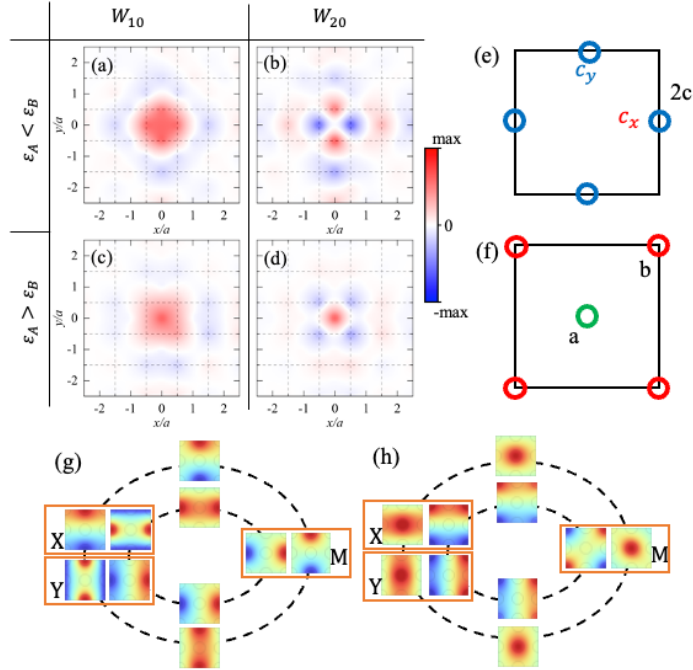

**Figure 3**. MLWFs obtained by FEM and the corresponding Wannier centers at maximal Wyckoff positions. (a)-(d) MLWF $W_{n0}(\mathbf{r})$ of the lowest two bands ($n = 1$ or $2$) for the 2D photonic crystal shown in Fig. 2(a): (a)-(b) $W_{n\mathbf{R}}(\mathbf{r})$ for the case of $\varepsilon_A < \varepsilon_b$ ($\varepsilon_A = 1$, $\varepsilon_B = 9.8$), (c)-(d) $W_{n\mathbf{R}}(\mathbf{r})$ for the case of $\varepsilon_A > \varepsilon_b$ ($\varepsilon_A = 9.8$, $\varepsilon_B = 1$). (e)-(f) Wannier centers within the unit cell, respectively, for $\varepsilon_A < \varepsilon_b$ and $\varepsilon_A > \varepsilon_b$. (g)-(h) The eigenstates evolution along the XM and YM for $\varepsilon_A < \varepsilon_b$ and $\varepsilon_A > \varepsilon_b$ which are the implement of Figures 1(d) and 1(g).

where $\eta$ denotes the parity as indicated by '+' and '−' in Figures 2c and 2d. The parities are in consistent with the symmetry-based analysis. The summation is over all the occupied bands $m$, and $i$ stands for $x$ or $y$. Due to the $C_4$ symmetry of the structure, we have $P_x = P_y$. Before and after the band inversion one obtains $\mathbf{P} = (1/2, 1/2)$ for the lowest two bands (the first and the second bands) by the mode parity. This is consistent with the results obtained by Wilson loop method (see Supporting Information S2).

Figures 3a-d show the real part of the MLWFs $W_{n\mathbf{0}}(\mathbf{r})$ ($n = 1, 2$) for these two lowest-energy bands that lie below the complete gap (see Supporting Information S3 for the calculation algorithm of MLWF and the complete complex $W_{n\mathbf{0}}(\mathbf{r})$). The imaginary part of $W_{n0}(r)$ nearly vanishes as the algorithm demands. It is seen that the four MLWFs $W_{n\mathbf{0}}(\mathbf{r})$ are all centered at the origin lattice ($\mathbf{R} = 0$) [31]. Intriguingly, their high localization occurs primarily at the unit cell center or edges, rather than in its interior. This contradicts with the conventional MLWFs and the nontrivial polarization obtained by equation (2). It is the band degeneracy that causes the individual local Wannier states ('orbitals') to be hybridized. Owing to the $C_4$ symmetry, the Wannier centers can be discerned at the centers of the unit's edges, denoted as 2c ($c_x$ and $c_y$) in Figure 3c. More specifically, for $\varepsilon_A < \varepsilon_B$ the in-phase ($|c_x\rangle + |c_y\rangle$) and out-of- phase ($|c_x\rangle - |c_y\rangle$) coupling between the local Wannier states $|c_x\rangle$ and $|c_y\rangle$ are represented by the calculated MLWFs in Figures 3a and 3b, respectively. In contrast to the Wannier functions which are usually localized inside the structure cell, these MLWFs are mainly centered at the cell edges as schematically plotted in Figure 3e. Similarly, Figure 3f illustrates the coupling results for the case of $\varepsilon_A > \varepsilon_B$, i.e., the in-phase $|a\rangle + |b\rangle$ and out-of-phase ($|a\rangle - |b\rangle$) hybridized Wannier states. The situations are also in consistent with the numerically obtained MLWFs [see Figures 3c and 3d]. In view of the highly localized characteristics of the WFs, they are suitable basis functions to establish tight-binding models which could approximate the system fairly well[32]. Alternatively, delving into the evolution of the eigenstates can further clarify the exchanging genesis of the Wannier bands ($v_x, v_y$).

Figures 3g and 3h depict the transformation of the eigenstates along XM and YM, in accord with the theoretical analysis presented in Figures 1d and 1g.

The configuration of Wannier states and their corresponding Wannier centers consistently manifest **P** = (1/2,1/2) for the two lowest bands, yet exhibiting quite different bulk states and corner charges (see Supporting Information S4). These characteristics, explored in a finite 10 × 10 sample enclosed by PEC boundaries, reveal specific distinctions: (1) the count of bulk states number rises from 180 for $\varepsilon_A > \varepsilon_B$ to 181 for $\varepsilon_A < \varepsilon_B$; (2) the edge states number for $\varepsilon_A > \varepsilon_B$ is 28, whereas it is 32 for $\varepsilon_A < \varepsilon_b$; (3) the corner charge (in normalized unit) decreases from 1.5 for $\varepsilon_A > \varepsilon_B$ to 1 for $\varepsilon_A < \varepsilon_B$. These findings can be validated by numerically computing the band structure and the density of states below the gap[31]. The observed distinctions in our sample—namely, the variation in bulk and edge states, along with the quantization of corner charge—highlight a nuanced interplay between material parameters and topological characteristics. Specifically, the subtle shift in bulk states and the quantifiable change in edge and corner states reflect the system's sensitivity to $\varepsilon_A$ and $\varepsilon_B$ values. These findings not only elucidate the influence of material parameters on topological properties but also set the stage for a deeper examination of the modal spectrum and the emergence of specific corner states within such a finite-sized photonic crystal. Figures 4a and 4b depict the modal spectrum of the finite-sized 10 × 10 sample for the cases of $\varepsilon_A > \varepsilon_B$ and $\varepsilon_A < \varepsilon_B$, respectively. These figures illustrate that both configurations can support edge (blue dots) and corner states (red dots), in addition to the bulk modes (gray dots). However, variations in corner state numbers and the modal profiles emerge due to the distinct MLWFs. More specifically, Figure 4c presents corner states **i**, **ii**, **iii** as marked in Figure 4a, distinguished by the first, second, and zero-order resonances characteristics, respectively[31]. Meanwhile, Figure 4d exhibits the corner states **iv** and **v** as marked in Figure 4b, defined by zero and first-order resonances respectively. It is pertinent to note that the spectrum of these distinct corner states could be modulated by adjusting the distance between the sample and surrounding PECs (see Supporting Information S5), enabling them to traverse across the bandgap and eventually merge into the continuum. Also, these corner states are pretty robust with respect to lattice disorder (see Supporting Information S6).

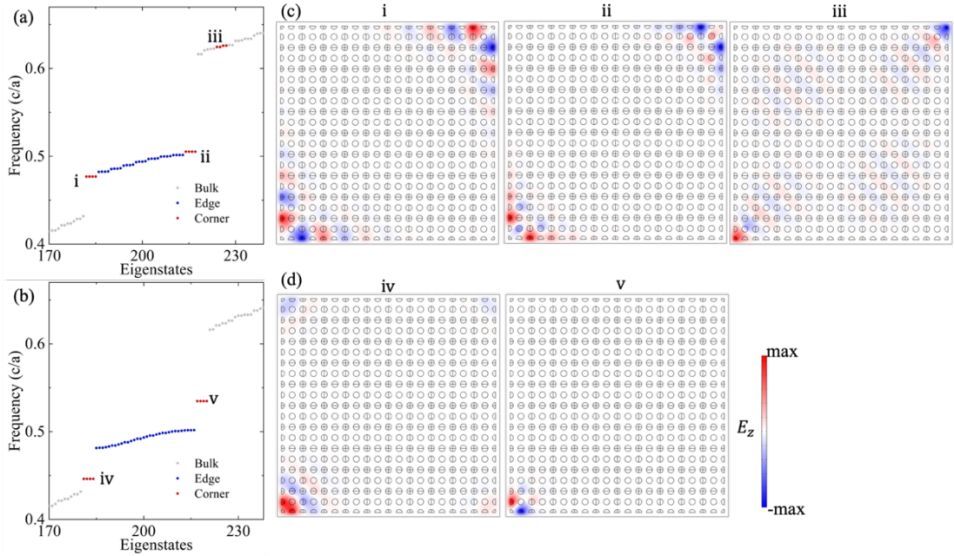

**Figure 4.** Modal spectra of finite-sized samples enclosed by PEC boundaries and the modal field distribution of the corner states. The resonant spectrum of the photonic crystal with (a) $\varepsilon_A = 9.8$, $\varepsilon_B = 1$ and (b) with $\varepsilon_A = 1$, $\varepsilon_B = 9.8$. (c) The field profiles of the corner states (**i, ii, iii**) labeled in (a), showing the first, second, and zero-order characteristic, respectively. (d) The field profiles of the corner states (**iv, v**) labeled in (b). The distance between the photonic crystal and PEC are $0.2a$. Note that $\varepsilon_{A,B} = 1$ suggests absence of the corresponding cylinders in the experimental samples.

**3. Experimentally measurements**

To experimentally observe and examine these distinct corner states in more depth, we fabricated two glide-symmetric photonic crystal using coupled $Al_2O_3$ cylindrical rods and conducted microwave-frequency measurements (see more details in Supporting Information S8). Experimentally, the quasi-2D photonic crystal, characterized by transverse-magnetic polarization, are covered by aluminium frames cladded on the top and bottom of the sample. These parallel metal plates are approximately 9 mm apart. The ceramic cylinders have uniform height of 8 mm and a radius of 2.5 mm. To avoid photon radiation into free space, we set up PEC, also represented by aluminium frames, around the exterior boundaries of the samples.

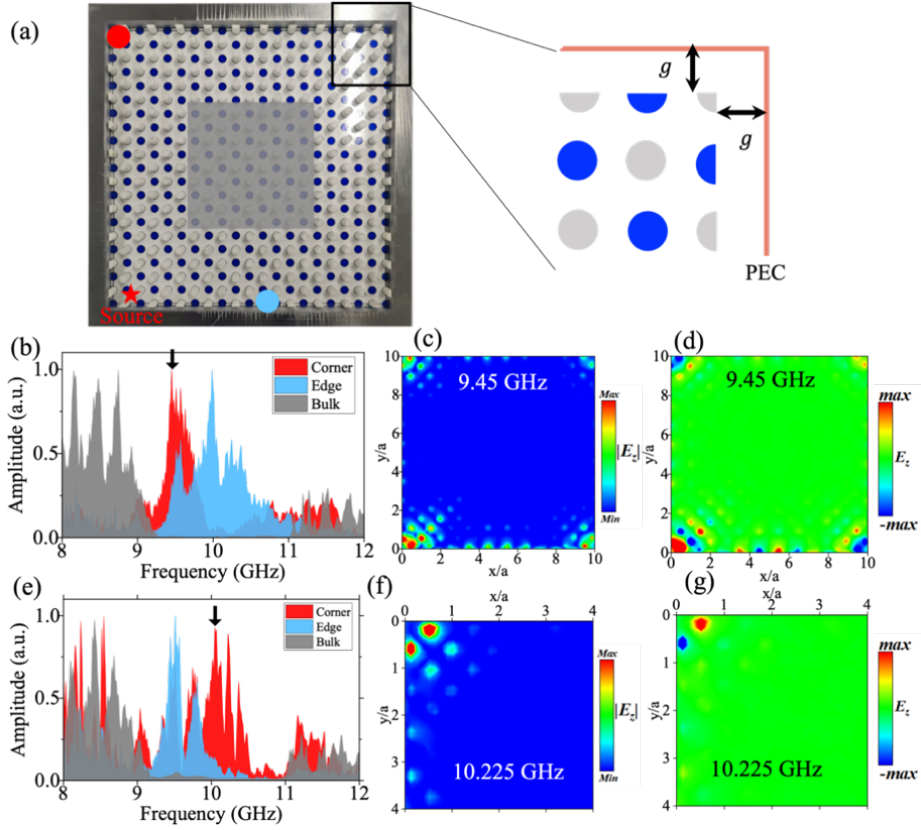

**Figure 5**. Experimental resonance spectral and modal pattern for sample A. (a) Top view of the dielectric cylinder array ($\varepsilon_A < \varepsilon_B$) cladded by the PEC with a gap $g$ and the schematic illustration of the corner structure. (b) Pump-probe spectroscopy with the source situated near the corner for the distance $g = 0.1a$. (c) and (d) Experimentally measured amplitude and the real part of electric field for the corner states marked by the downward black arrow in (b). (e) Pump-probe spectroscopy for the distance $g = 0.2a$. (g) and (g) The measured amplitude and the real part of the electric field for the corner state marked by the downward black arrow in (e).

We compare the two kinds of photonic crystal, one with $\varepsilon_A < \varepsilon_B$ (Sample A) and the other with $\varepsilon_A > \varepsilon_B$ (Sample B), to highlight the lattice symmetry-dependent higher-order topological features.

Figure 5a presents the top view of the sample A characterized by $\varepsilon_A < \varepsilon_B$. Figure 5b illustrates the 'pump-probe'-like spectroscopy with the source sitting near the corner, denoted by the red star (see Supporting Information S8 for more details). Note that the probe-received signal amplitude depends on its position, as shown by the dots at the corner and edge in Figure 5a that are meant for detecting the corner states and the edge states, respectively. The bulk modes are identified when the probe is placed inside the bulk region indicated by the shaded square in Figure 5a. Figure 5b clearly shows that a corner state resonance emerges at $f = 9.45$ GHz (marked by the downward black arrow) within the shared spectral gap. The microwave excitation inside the sample was then mapped by setting the pump exactly at one specific frequency [33]. As the working frequency increases, bulk mode gradually evolves into localized ones at the corners (Figure 5c), and the corner states then starts to couple with the edge states, ultimately turning into a bulk state again. To evaluate the modal symmetry properties, the real part electric field of the corner state is also measured, as exhibited in Figure 5d for the corner state at $f = 9.45$ GHz. It is apparent that the state is symmetric with respect to the sample's diagonal line, in accord to the numerically calculated state **v** in Figure 4d. In this scenario, the accompanying state **iv** becomes unobservable due to its immersion inside the bulk modes. To render it observable, the distance between the sample and the enclosed PEC is increased from $0.1a$ to $0.2a$. The evolution of the pump-probe spectroscopy in this modified arrangement is shown in Figure 5e. An additional corner state appears in the spectrum at $f = 10.225$ GHz that lies exactly within the bulk band gap, indicated by the black arrow in Figure 5e. This corner state exhibits anti-symmetric feature, clearly reflected in the measured intensity and electric field profiles in Figures 5f and 5g, respectively.

To this end, we have actually verified the situation in Figures 3a, 3b and 3e. We have also fabricated samples (Sample B) according to the cases in Figures 3c, 3d and 3f. Figure 6(a) displays the top view of the sample made by $Al_2O_3$ rods again. For this structure we have unit cell defined in Figure 2 that obeys $\varepsilon_A > \varepsilon_B$. In this instance, two corner states emerge inside the spectral gap, as indicated by the downward black arrows in Figure 6b: one at $f = 9.135$ GHz, exhibiting characteristics akin to corner state **i** in Figure 4c, and the other at $f = 9.985$ GHz, resembling corner state **ii** in Figure 4c. Figures 6c-f show the corresponding measured electric field intensity distribution and profiles which reveal a clear indication of corner localized wave excitation of the topological corner states. To support corner state **iii** as described in Figure 4c, the gap between the sample and the PEC is expanded from $0.2a$ to $0.3a$. The evolution of the pump-probe spectroscopy within this altered boundary setup is shown in Figure 6g. As marked by the downward arrow, an additional resonant state, embedded within the bulk band gap is identified. This corner state demonstrates pronounced field localization, a feature distinctly reflected in the measured electric field profiles in Figures 6h and 6i. As expected, the modal field symmetry is identical to that in Figure 4c.

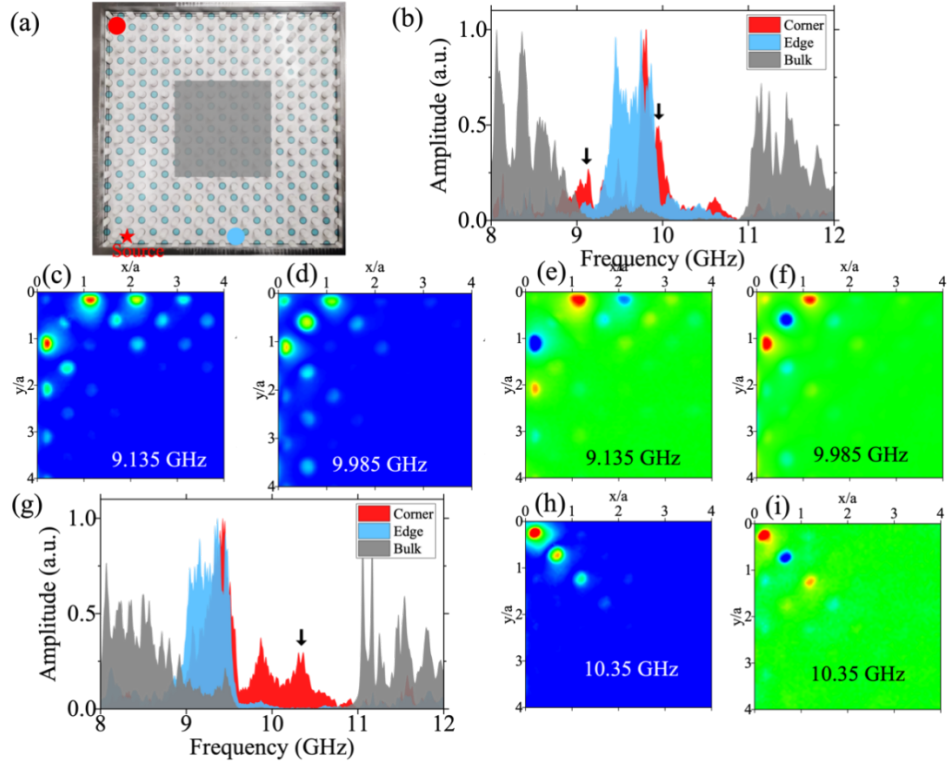

**Figure 6.** Experimental resonance spectral and modal pattern for sample B. (a) Top view of the dielectric cylinder array ($\varepsilon_A > \varepsilon_B$) cladded by the PEC. (b) Pump-probe spectroscopy with the source situated near the corner for the distance $g = 0.1a$. (c) and (d) Experimentally measured electric field profile $|E_z|$ for corner states. (e) and (f) The corresponding real part of electric field for (c) and (d), respectively. (g) Pump-probe spectroscopy for the distance $g = 0.2a$. (h) and (i) The measured amplitude and the real part of the electric field for the corner state indicated in (g).

## 4. Conclusions

To conclude, we propose a case of Wannier band inverting in orthogonal directions which enables distinct topological phase in photonic crystal with glide symmetry. It is shown that the bulk polarization that are usually adopted in a summation fashion fails to describe the band topology in some cases with partially degenerated bands. Topologically distinct phases with equal polarization vector $(P_x, P_y)$ are identified and the Wannier bands are examined in a higher-order form by tracing the spatial distribution and symmetry of the associated maximally localized Wannier functions optimized for two degenerated bands with appropriate gauge. It is expected that involving dislocations in the photonic crystal studied here could yield topological defect states as well, guaranteed by the fractional filling anomaly[34]. We argue that the Wannier function concept borrowed from electronic systems shall be carefully adopted for photonic systems and could play important role in studying topological phases and reveal new topological physics in bosonic systems. Studies involving simple symmetry identities at high momentum points shall be examined in much care. In this regard, it is interesting to note that recently Xu et al showed that although being topologically distinct, the well-known Wu-Hu model offering two phases that had long been believed possessing spin Hall phases are essentially trivial, in the sense that symmetric localized Wannier functions do exist (not obstructed). It was thus argued that both phases by shrunken and expanded lattice do not enforce any topological protection to the interface states[35]. Our results presented here highlight an exceptional example on the importance of tracing exactly the Wannier function evolution across all the relevant energy bands and the significance of band-resolved polarization constituents.

**Supporting Information**
Supporting Information is available from the Wiley Online Library or from the author.

**Acknowledgements.**

This work was supported by the National Natural Science Foundation of China (Nos. 62375064, 92050104, 12274314, and 12074279), Shenzhen Science and Technology Program (No. JCYJ20210324132416040), Guangdong Provincial Nature Science Foundation (No. 2022A1515011488, 2023A1515110572), the Major Program of Natural Science Research of Jiangsu Higher Education Institutions (No.18KJA140003), and the National Key Research and Development Program of China (No. 2022YFB3603204).

**Author contributions**

Z. L and J. X conceived the idea. Z. L did the theoretical calculation, designed the samples and performed the simulations. X. Z performed the experimental measurement. J. X supervised the project. Z. L and J. X. wrote the manuscript. All authors contributed to scientific discussions of the manuscript.

**Conflict of interest**

The authors declare no competing interests.

**Data Availability Statement**

The data that support the findings of this study are available from the corresponding author upon reasonable request.

Supporting Information

# Topological Corner Modes by Composite Wannier States in Glide-Symmetric Photonic Crystal

*Zhenzhen Liu[1,2,#], Xiaoxi Zho[3,4,#], Guochao Wei[5], Lei Gao[3,4], Bo Hou[6], and Jun-Jun Xiao[1,*]*

[1]Shenzhen Engineering Laboratory of Aerospace Detection and Imaging, College of Electronic and Information Engineering, Harbin Institute of Technology (Shenzhen), Shenzhen 518055, China.

[2]College of Science, Shantou University, Shantou 515063, China.

[3]School of Optical and Electronic Information, Suzhou City University, Suzhou 215104, China

[4]School of Physical Science and Technology, Soochow University, Suzhou 215006, China

[5]School of Mathematical and Physical Sciences, Wuhan Textile University, Wuhan 430200, China

[6]Wave Functional Metamaterial Research Facility, The Hong Kong University of Science and Technology (Guangzhou), Guangzhou 511400, China

* E-mail: eiexiao@hit.edu.cn

[#]These two authors contributed equally.

# Table of contents

## S1. Symmetry-based analysis and the parity at high-symmetric points

The two-dimensional (2D) photonic crystals (PCs) consist of cylinders with center-to-center distance $a/2$ as shown in Fig. 2(a). The choice of the enlarged unit cell outlined by the black-dashed square is in fact a doubling of the primitive cell. Clearly, the primitive unit cell holds the glide symmetry $G_x: (x,y) \rightarrow (-x+\frac{a}{2}, y+\frac{a}{2})$ and $G_y: (x,y) \rightarrow (x+\frac{a}{2}, -y+\frac{a}{2})$, the fourfold rotational symmetry $C_4: (x,y) \rightarrow (y,x)$, the translation symmetry $\tau: (x,y) \rightarrow (x+a/2, y+a/2)$, the inversion symmetry $I: (x,y) \rightarrow (-x,-y)$, the mirror symmetry $M: (x,y) \rightarrow (-x,y)$, and the time reversal symmetry $T$. The anti-unitary symmetry operators $\Theta_i = G_i T$ ($i = x, y$) enable double degeneracy at the whole edges of the Brillouin zone (BZ) due to the Kramer's double degeneracy [1].

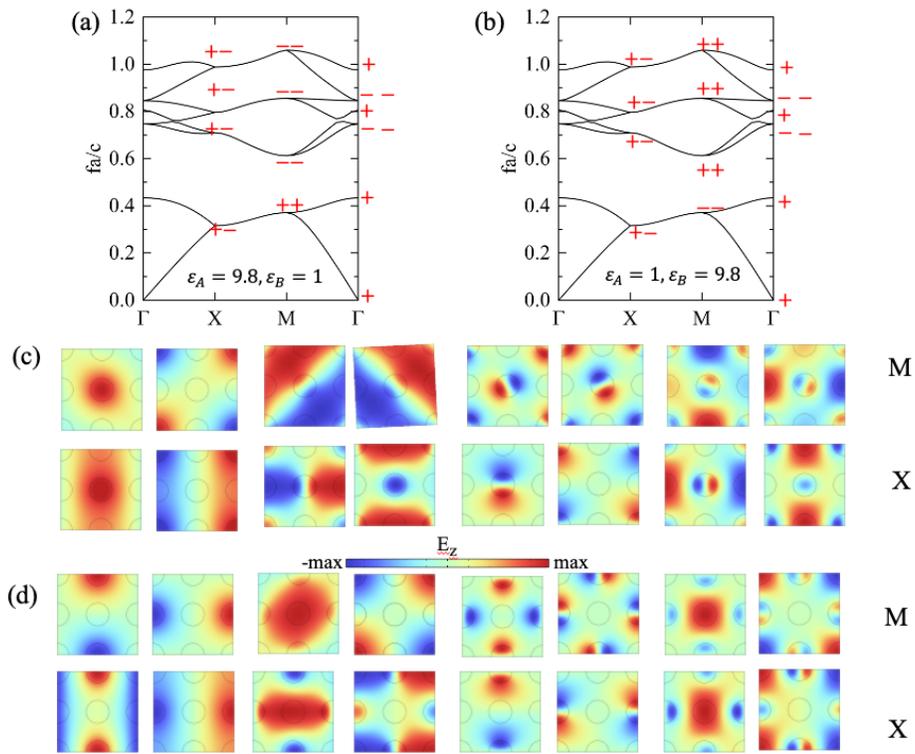

**Figure. S1.** Photonic band structures of the PCs shown in Fig. 2(a) for different permittivity parameters (a) $\varepsilon_A = 9.8$, $\varepsilon_B = 1$ and (b) $\varepsilon_A = 1$, $\varepsilon_B = 9.8$, where "+" and "-" indicates the even and odd parity of the photonic eigenstates for inversion operation, respectively. The electric fields profiles $E_z$ at M and X for (a) and (b), respectively, are presented in (c) and (d).

In particular, we have $\Theta_i I = -I\Theta_i$ at the X and Y points of the first Brillouin zone, whereas $\Theta_i I = I\Theta_i$ holds at M point. Therefore, the doubly degenerated eigenstates $\psi_{n,k}$ and $\Theta_i\psi_{n,k}$ have opposite parities for the inversion operation $I$ at X and Y points. On the contrary, they have identical parities for the inversion operation $I$ at M point. These can be verified by the band structure and the field distribution shown in Figs. S1 and Fig. S2 for $\varepsilon_B = 1$ and $\varepsilon_B = 4$. The parities are indicated by "+" and '−" and are consistent with the prediction of the symmetry operations. Clearly, before and after the topological phase transition by the gap opening, bands below the photonic bandgap hosts the total polarization $\mathbf{P} = (\frac{1}{2}, \frac{1}{2})$, implying the existence of nontrivial topological phases.

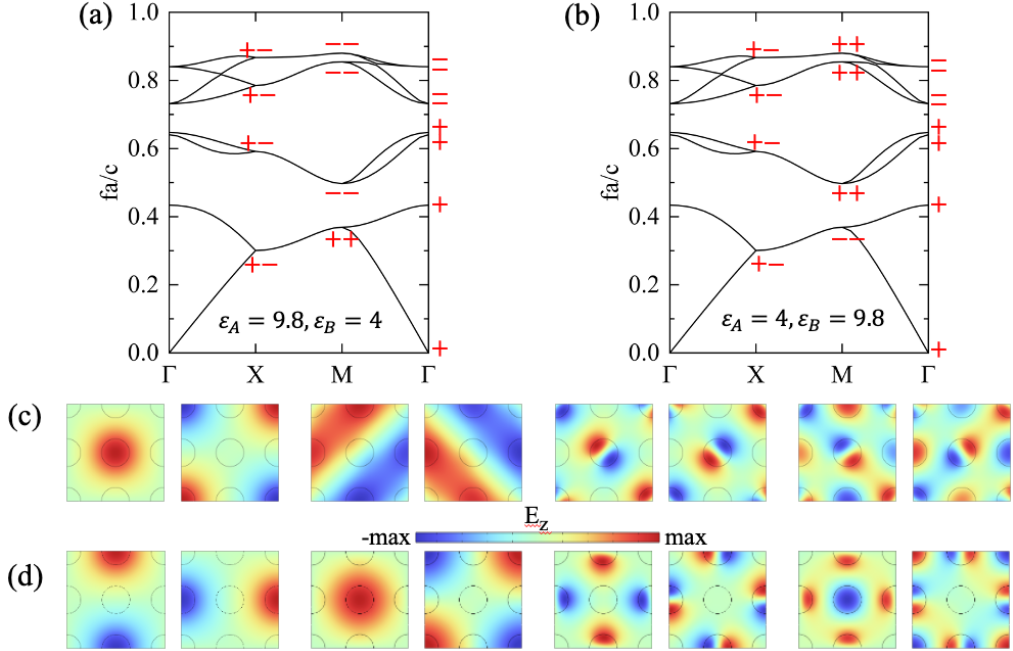

**Figure. S2.** Similar to Fig. S1 but for the case that (a) $\varepsilon_A = 9.8$, $\varepsilon_B = 4$ and (b) $\varepsilon_A = 4$, $\varepsilon_B = 9.8$. The modal patterns for M points are shown in (c) and (d), respectively.

## S2. Numerical obtained Wannier bands from Wilson loop method

The calculations of Wannier band $v_y$ and $v_x$ can be obtained by Wilson loop along the x and y directions which is constructed as $W_{x(y),\mathbf{k}} = F_{x(y),\mathbf{k}+N_{x(y)}\Delta k_{x(y)}} \ldots F_{x(y),\mathbf{k}+\Delta k_{x(y)}} F_{x(y),\mathbf{k}}$ with the starting point of the closed loop $k = (k_x, k_y)$. Here, $[F_{x,\mathbf{k}}]^{m,n}$ is an element of matrix $F_{x,\mathbf{k}}$, an inner product of Bloch eigenstates at adjacent k point. The eigenvalues of the Wilson loop operator is referred as the Wannier band as shown in Fig. **S3**. However, the two Wannier bands cannot be directly combined to for two polarizations $\mathbf{P}_1$ and $\mathbf{P}_2$. As such, the Wannier functions is used to identify the composed polarizations, detailed in the following section.

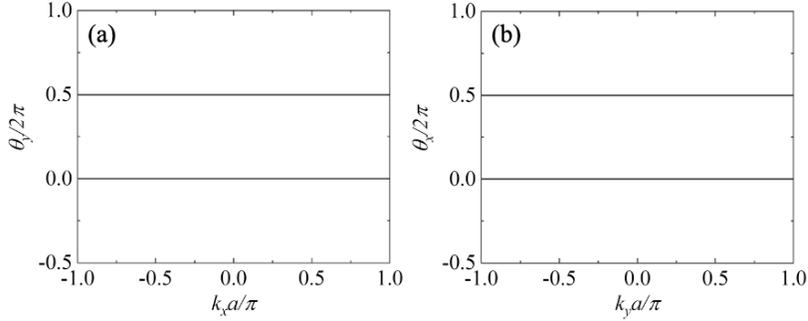

**Figure. S3.** Wannier band, (a) $v_y = \theta_y/2\pi$ and (b) $v_x = \theta_x/2\pi$, for the lowest two energy bands of the PCs: $\varepsilon_A < \varepsilon_B$ and $\varepsilon_A > \varepsilon_B$.

## S3. Maximally localized Wannier functions

We basically follow the method in Refs. [2,3] to calculate the Wannier functions, starting from the electromagnetic wave equation in terms of the electric field

$$\nabla \times (\mu^{-1} \nabla \times \mathbf{E}) = \left(\frac{\omega}{c}\right)^2 \varepsilon(\mathbf{r})\mathbf{E}, \tag{S1}$$

where $\varepsilon(\mathbf{r})$ is the spatially varying relative dielectric function and $\mu = 1$. The Wannier functions are obtained from the Bloch functions, $\mathbf{u}_{n\mathbf{k}}(\mathbf{r})$, in the case of a periodic structure with lattice constant $a$. The Wannier functions in two dimensions are defined by the transformation

$$w_n(\mathbf{r} - \mathbf{R}) = \frac{A}{(2\pi)^2} \int_{BZ} d\mathbf{k} e^{-i\mathbf{k}\cdot\mathbf{R}} \mathbf{u}_{n\mathbf{k}}(\mathbf{r}), \tag{S2}$$

where A is the area of the real space unit cell.

For crossing bands, it is conventional to replace the Wannier functions for the individual bands in the group by a set of generalized Wannier functions. The generalized Wannier functions are related to the Bloch functions by

$$W_{nR}(\mathbf{r} - \mathbf{R}) = \frac{A}{(2\pi)^2} \int_{BZ} d\mathbf{k} e^{-i\mathbf{k}\cdot\mathbf{R}} \sum_m U_{mn}^{(\mathbf{k})} \mathbf{u}_{n\mathbf{k}}(\mathbf{r}), \tag{S3}$$

where $U_{mn}^{(\mathbf{k})}$ is an $N \times N$ unitary matrix which mixes the N bands in the group at wave vector $\mathbf{k}$. Hence, the problem of choosing the gauge-free phases to minimize the spread of Eq. (S3) becomes one of finding the set of unitary transformations that minimizes the average spread of the functions in the group.

To compute the maximally Wannier function in our case, the aim is to find the set of unitary matrices $U_{mn}^{(\mathbf{k})}$ on a grid of $21 \times 21$ points spanning the Brillouin zone, which minimizes the spread function

$$\Omega = \sum_n [\langle r^2 \rangle_n - \langle \mathbf{r} \rangle_n^2], \tag{S4}$$

where $\langle \mathbf{r} \rangle_n^2 = \frac{iV_c}{(2\pi)^2} \int \langle u_{n\mathbf{k}}(\mathbf{r}) | \partial_k u_{n\mathbf{k}}(\mathbf{r}) \rangle d\mathbf{k}$ and $\langle r^2 \rangle_n = \frac{iV_c}{(2\pi)^2} \int \langle |\partial_k u_{n\mathbf{k}}(\mathbf{r})|.|\partial_k u_{n\mathbf{k}}(\mathbf{r}) \rangle d\mathbf{k}$.

Finally, Eq. (S4) can be expressed in terms of a set of overlap matrices

$$M_{mn}^{(\mathbf{k},\mathbf{b})} = \langle u_{z,m,\mathbf{k}}(\mathbf{r}) | u_{z,m,\mathbf{k}+\mathbf{b}}(\mathbf{r}) \rangle, \tag{S5}$$

where $u_{z,m,\mathbf{k}}(\mathbf{r})$ and $u_{z,m,\mathbf{k}+\mathbf{b}}(\mathbf{r})$ are the periodic parts of the Bloch function solution to Eq. (S2) at grid point $\mathbf{k}$ and its near neighbor $\mathbf{k}+\mathbf{b}$. To obtain the minimum of Eq. (S5), the steepest descent method is used. The gradient of the spread function at each grid $\mathbf{k}$ is

$$G^{(\mathbf{k})} = \frac{d\Omega}{dW^{(\mathbf{k})}} = 4 \sum_\mathbf{b} w_b \left( \mathcal{A}[R^{(\mathbf{k},\mathbf{b})}] - \mathcal{S}[T^{(\mathbf{k},\mathbf{b})}] \right), \tag{S6}$$

where $\mathcal{A}$ and $\mathcal{S}$ are the superoperators $\mathcal{A}[B] = (B - B^\dagger)/2$ and $\mathcal{S}[B] = (B + B^\dagger)/2i$, $R_{mn}^{(\mathbf{k},\mathbf{b})} = M_{mn}^{(\mathbf{k},\mathbf{b})} M_{nn}^{(\mathbf{k},\mathbf{b})*}$, $T_{mn}^{(\mathbf{k},\mathbf{b})} = \tilde{R}_{mn}^{(\mathbf{k},\mathbf{b})} q_n^{(\mathbf{k},\mathbf{b})}$, $\tilde{R}_{mn}^{(\mathbf{k},\mathbf{b})} = M_{mn}^{(\mathbf{k},\mathbf{b})}/M_{nn}^{(\mathbf{k},\mathbf{b})}$, $q_{nn}^{(\mathbf{k},\mathbf{b})} = \text{Im} \ln M_{nn}^{(\mathbf{k},\mathbf{b})} + \mathbf{b}\cdot\mathbf{r}_n$.

The condition for having found a minimum is that $G^{(\mathbf{k})}$ in Eq. (S6) vanishes. Then one repeats the cycle to update $U_{mn}^{(\mathbf{k})}$ until convergence is obtained. Following the aforementioned steps, we have found the maximally localized Wannier functions (MLWFs). Figures **S4** and **S5** show the real and imaginary parts of the MLWFs for the two lowest energy bands. Clearly the imaginary parts of the obtained MLWFs are extremely small, as such they can be considered to be real.

Besides, we also calculate the general PCs consisting of four equal rods. There exist two different configurations to break the degeneracy at M by moving the rods towards the center or far away from the center, as illustrated in the top of Fig. **S6**. In this case, the lowest band is separated with others. Using the aforementioned method, the MLWFs can also be obtained, whose corresponding centers are located at (0,0) and (1/2,1/2).

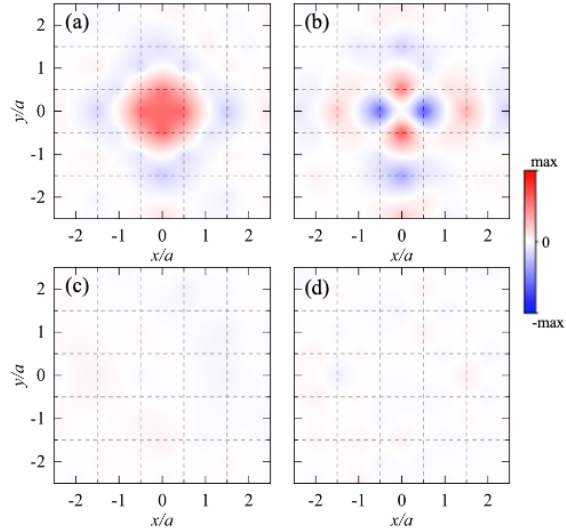

**Figure. S4.** The MLWFs for PCs with $\varepsilon_A < \varepsilon_B$. (a) and (b) The real part. (c) and (d) The imaginary part.

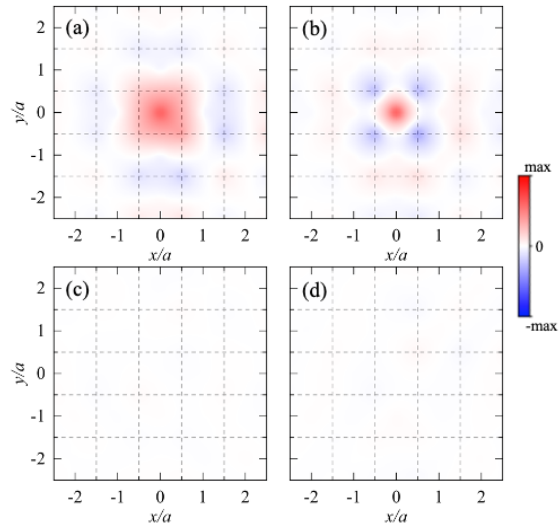

**Figure. S5**. The MLWFs for PCs with $\varepsilon_A > \varepsilon_B$. (a) and (b) The real part. (c) and (d) The imaginary part.

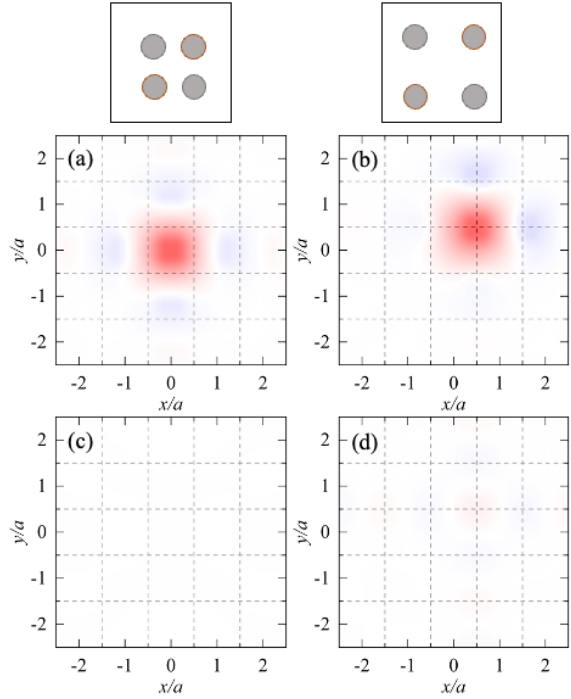

**Figure. S6**. The MLWFs for two typical PCs with $\varepsilon_A = \varepsilon_B = 9.8$. (a) and (c) The constituent four rods of the unit cell move towards the center. (b) and (d) The constituent four rods of the unit cell move towards the corner. (a) and (b) The real part. (c) and (d) The imaginary part.

**S4. Fractional charge calculated for the finite-size samples**

The number of bulk states ("spectral charge") contributed by the corner, edge and bulk unit cells can be calculated by the local density of states (LDOS). The spectral charge for the $i$-th unit cell in the finite structure is defined as [4,5]

$$Q_i = \int_0^{f_{gap}} df \int_{i-U.C} d\mathbf{r}\, \rho(f,\mathbf{r}), \tag{S7}$$

where the integration is over the $i$-th unit cell, $f_{gap}$ is a frequency in the complete band gap of the bulk and edge states, and $\rho(f,\mathbf{r})$ is the LDOS of photons obtained through a summation weighted by the amplitude of the eigenstates at point $\mathbf{r}$ [6]:

$$\rho(f,\mathbf{r}) = \sum_i |\Psi_i(\mathbf{r})|^2 \delta(f - f_i), \tag{S8}$$

where $\Psi_i$ denotes the $i$-th photonic eigenstate with frequency $f_i$. In our calculation, the normalized photonic eigenstate is given by

$$|\Psi_i(\mathbf{r})|^2 = \varepsilon(\mathbf{r})|E_{z,i}(\mathbf{r})|^2, \tag{S9}$$

where $\varepsilon(\mathbf{r})$ is the relative permittivity and $E_{z,i}(\mathbf{r})$ is the rescaled electric field of the $i$-th photonic eigenstate that satisfies $\int d\mathbf{r}\varepsilon(\mathbf{r})|E_{z,i}(\mathbf{r})|^2 = 1$.

The spectral charge at the $i$-th unit cell is then obtained by Eq. (S7), where the integration over the coordinates is performed for the region inside the $i$-th unit cell. Figure S7 shows the obtained charge distribution for two distinct phases: $\varepsilon_A > \varepsilon_B$ and $\varepsilon_A < \varepsilon_B$.

(a)

| 1.25 | 1.50 | 1.50 | 1.50 | 1.50 | 1.50 | 1.50 | 1.50 | 1.50 | 1.25 |
|---|---|---|---|---|---|---|---|---|---|
| 1.50 | 2.00 | 2.00 | 2.00 | 2.00 | 2.00 | 2.00 | 2.00 | 2.00 | 1.50 |
| 1.50 | 2.00 | 2.00 | 2.00 | 2.00 | 2.00 | 2.00 | 2.00 | 2.00 | 1.50 |
| 1.50 | 2.00 | 2.00 | 2.00 | 2.00 | 2.00 | 2.00 | 2.00 | 2.00 | 1.50 |
| 1.50 | 2.00 | 2.00 | 2.00 | 2.00 | 2.00 | 2.00 | 2.00 | 2.00 | 1.50 |
| 1.50 | 2.00 | 2.00 | 2.00 | 2.00 | 2.00 | 2.00 | 2.00 | 2.00 | 1.50 |
| 1.50 | 2.00 | 2.00 | 2.00 | 2.00 | 2.00 | 2.00 | 2.00 | 2.00 | 1.50 |
| 1.50 | 2.00 | 2.00 | 2.00 | 2.00 | 2.00 | 2.00 | 2.00 | 2.00 | 1.50 |
| 1.50 | 2.00 | 2.00 | 2.00 | 2.00 | 2.00 | 2.00 | 2.00 | 2.00 | 1.50 |
| 1.25 | 1.50 | 1.50 | 1.50 | 1.50 | 1.50 | 1.50 | 1.50 | 1.50 | 1.25 |

(b)

| 1.02 | 1.49 | 1.50 | 1.50 | 1.50 | 1.50 | 1.50 | 1.50 | 1.49 | 1.02 |
|---|---|---|---|---|---|---|---|---|---|
| 1.49 | 1.99 | 2.00 | 2.00 | 2.00 | 2.00 | 2.00 | 2.00 | 1.99 | 1.49 |
| 1.50 | 2.00 | 2.00 | 2.00 | 2.00 | 2.00 | 2.00 | 2.00 | 2.00 | 1.50 |
| 1.50 | 2.00 | 2.00 | 2.00 | 2.00 | 2.00 | 2.00 | 2.00 | 2.00 | 1.50 |
| 1.50 | 2.00 | 2.00 | 2.00 | 2.00 | 2.00 | 2.00 | 2.00 | 2.00 | 1.50 |
| 1.50 | 2.00 | 2.00 | 2.00 | 2.00 | 2.00 | 2.00 | 2.00 | 2.00 | 1.50 |
| 1.50 | 2.00 | 2.00 | 2.00 | 2.00 | 2.00 | 2.00 | 2.00 | 2.00 | 1.50 |
| 1.50 | 2.00 | 2.00 | 2.00 | 2.00 | 2.00 | 2.00 | 2.00 | 2.00 | 1.50 |
| 1.49 | 1.99 | 2.00 | 2.00 | 2.00 | 2.00 | 2.00 | 2.00 | 1.99 | 1.49 |
| 1.02 | 1.49 | 1.50 | 1.50 | 1.50 | 1.50 | 1.50 | 1.50 | 1.49 | 1.02 |

**Figure S7**. The charge distributions of the finite samples ($10 \times 10$) enclosed by PEC boundaries. (a) $\varepsilon_A = 9.8 > \varepsilon_B = 1$, (b) $\varepsilon_A = 1 < \varepsilon_B = 9.8$.

### S5. Spectral evolution of the corner states

It has been realized that the corner states exist in both the band gap of the lowest two bands for $\varepsilon_A > \varepsilon_B$ and $\varepsilon_A < \varepsilon_B$. They can generally couple and hybridize with the bulk and/or edge modes. The corner states can be spectrally manipulated by tuning the relative permittivity between $\varepsilon_A$, $\varepsilon_B$, or by tuning the gap $\Delta_x$ and $\Delta_y$ between the finite sample and the perfect electric conductor (PEC). Figures **S8** shows the corner mode dispersion for the corner between a finite PCs enclosed by PEC with gap $\Delta_x = \Delta_y = 0.2a$. Clearly, the corner states merge into the lower or higher bulk band when the permittivity $\varepsilon_A$ or $\varepsilon_B$ approaches to 1. On the other hand, when the gap $\Delta_x$ and $\Delta_y$ gradually increase, the corner states can also be embedded into the bulk states, as shown in Fig. **S9**.

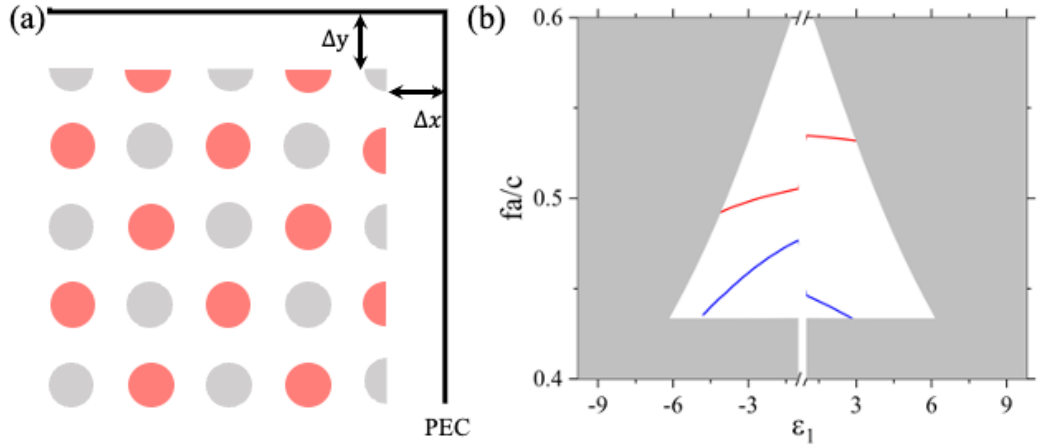

**Figure. S8**. (a) The schematic of the PCs structure enclosed by PEC with a small gap $\Delta_x$, $\Delta_y$. (b) The evolution of bulk and corner bands are illustrated as the function of relative permittivity $\varepsilon_I$: $\varepsilon_I = \varepsilon_B$ when $\varepsilon_A = 9.8$, $\varepsilon_I = -\varepsilon_A$ when $\varepsilon_B = 9.8$.

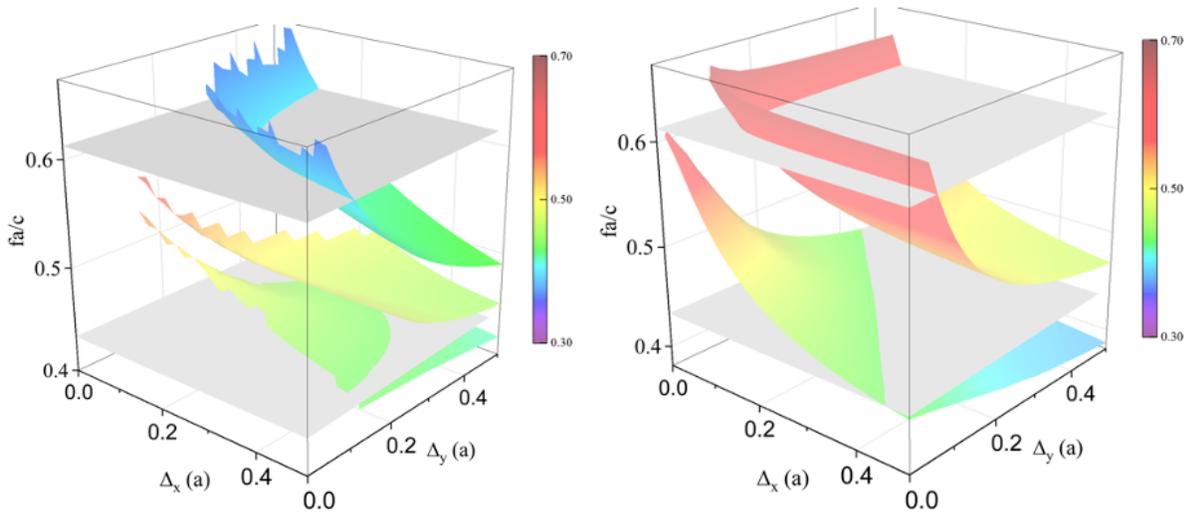

**Figure. S9**. The evolution of bulk and corner bands are illustrated as the function of $\Delta_x$ and $\Delta_y$ for the case of (a) $\varepsilon_A = 9.8$, $\varepsilon_B = 1$, (b) $\varepsilon_A = 1$, $\varepsilon_B = 9.8$. The gray planes correspond to the edge of the bulk.

## S6. Robustness of the corner states against disorder around the corner

As explored in the main text, corner states, protected by nontrivial bulk polarization and hybridized Wannier functions, are robust to certain levels of disorder due to their glide symmetry ensured topological protection. This can be validated through numerical simulations introducing disorder around the corners where corner states primarily localize. Figure **S10a** illustrates the band structure of a 10×10 unit cell, enclosed by PEC. Here, rods at the bottom-left corner shift towards the center by approximately 0.05a, as shown in inset of Fig. **S10c**. Despite the introduced disorder, different corner states persist. Compared to Fig. **4(a)**, the frequency and field profile of corner states **i** and **ii** remain largely unchanged. Conversely, corner state **iii** transitions from the bulk to the gap due to the displacement of its resonant cavity from the PEC. In the scenario of Fig. **S10b**, the frequency of corner states **iv** and **v**, primarily localized at the shifted rods, diverges from the frequencies of other corner states localizing at different corners. This is attributed to the relocation of the corner states primarily localized at the shifted rods.

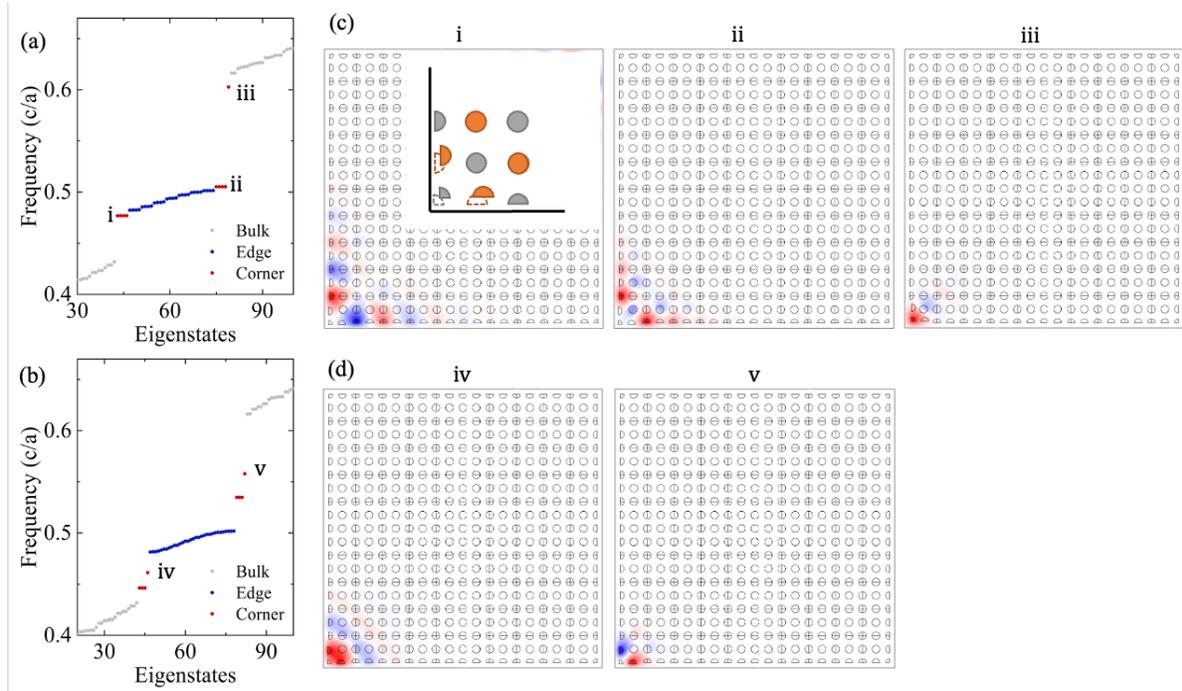

**Figure. S10**. Similar to Fig. 4 of the main text. The field profile of the corner states mainly localized in the vicinity of the left-bottom corner where the movement of the rods are moved away from bulk with distance $0.05a$.

## S7. Corner states in the box-shaped PC with different topological phases

The topology of the PCs, featured by the polarization $\mathbf{P} = (\frac{1}{2}, \frac{1}{2})$, has been clearly explored in the main text. Intuitively, the corner constructed by the PCs with $\varepsilon_A = 9.8$, $\varepsilon_B = 1$ surrounded by PCs with $\varepsilon_A = 1$, $\varepsilon_B = 9.8$ does not support a corner state due to their apparently identical polarization. However, Fig. **S11** shows that two corner states emerge due to their different Wannier functions and the resultant non-equal corner charge. This further manifests the higher-order topology enabled by exchanged Wannier band features.

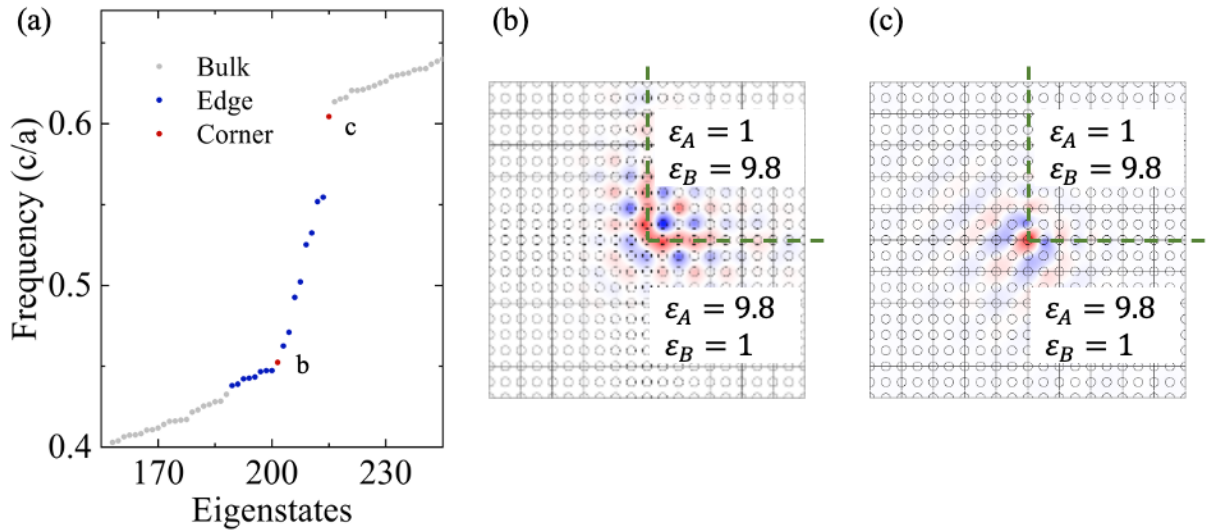

**Figure. S11**. Modal spectra of finite-sized samples enclosed by PEC boundaries and the modal field distribution of the corner states. (a) The resonant spectrum of the PCs with $\varepsilon_A = 1$, $\varepsilon_B = 9.8$ enclosed by PCs with $\varepsilon_A = 9.8$, $\varepsilon_B = 1$. (b)-(c) The field profiles of the corner states (b, c) labeled in (a). The black dashed curves indicate the interface between two PCs.

## S8. Microwave experimental measurement setup

The experimental system is based on quasi-2D photonic systems of transverse-magnetic polarization (electric field perpendicular to the 2D plane) created by metal plates cladded from above and below, as shown in **Fig. S12(a)**. To avoid friction between the sample and the plate, there was a 1mm air gap between the upper metal plate and the sample for sample movement. Samples are constructed by carefully pasting dielectric cylinders into a pre-printed drawing on which the location of each cylinder is marked. The dielectric cylinders and half-cylinders are made of commercial alumina ceramics (Al$_2$O$_3$) doped with chromium dioxide. These cylinders and half-cylinders have the same height of 8 mm and the same radius of 2.5 mm. The measurement instrument used was the N5230C vector network analyzer produced by Agilent Technologies (100MHz-20GHz). Port 1 of the analyzer was connected to a monopole antenna inserted into the sample through the lower metal plate as the excitation source, positioned approximately 6mm high. The probing source consisted of a coaxial cable with its outer metal layer stripped off, exposing a length of approximately 5mm, inserted into the air gap through the upper metal plate and connected to Port 2. By fixing the upper metal plate and moving the lower metal plate (**Fig. S12b**), we can obtain the S-parameters of each point in the sample. For the scanning measurement, the scanning step size was set to 2.5 mm, which is half the diameter of the sample, allowing for the effective measurement of the amplitude and phase of each cylindrical object.

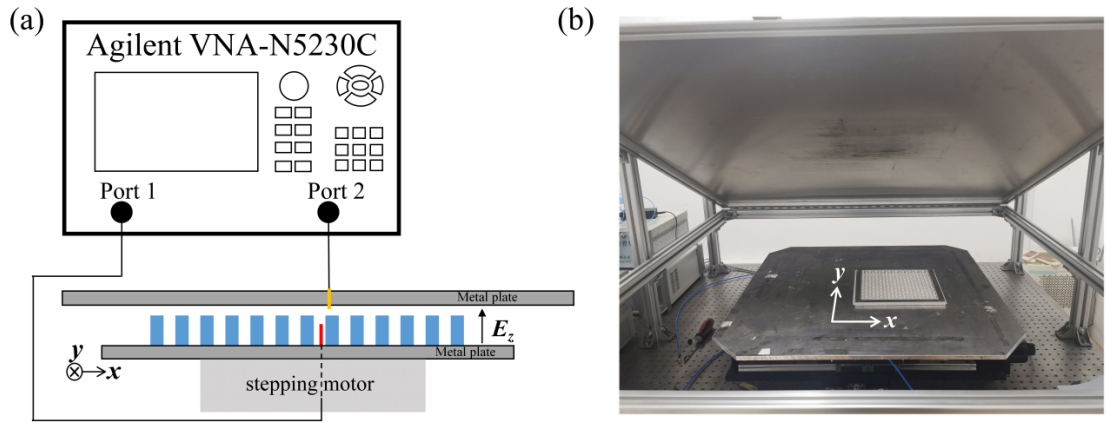

**Figure. S12. Experimental Set-up.** (a) Schematic diagram of experiment. The Deep gray area represents two parallel metal-slabs (as PEC) and the light gray is steeping motor, which confines transverse-magnetic harmonic modes in quasi-2D space. The red line and orange line represent the excitation antenna and the detection source, respectively. (b) Photograph of the set-up. The sample of aluminum oxide was placed at the center region between two-metal plates (the blue position in (**a**), which is encased in a square aluminum frame that has been cut by laser on all sides.

**Supplementary References**